\documentstyle[12pt,aaspp4,epsf]{article}

\received{}
\accepted{}
\journalid{}{}
\articleid{}{}

\slugcomment{\small \em To appear in ApJ, Vol. 
490, November 20, 1997}

\lefthead{Zwaan et al.}
\righthead{The H{\sc i} Mass Function}

\begin{document}

\newcommand{\hi}{\ion{H}{1} }
\newcommand{\hei}{\ion{He}{1} }
\newcommand{\mhi}{M_{\rm HI}}
\newcommand{\mhis}{M^*_{\rm HI}}
\newcommand{\msol}{{\rm M}_\odot}
\newcommand{\ohi}{\Omega_{\rm HI}}
\newcommand{\og}{\Omega_{\rm g}}
\newcommand{\rhi}{\rho_{\rm HI}}
\newcommand{\ml}{{\cal M}/L}
\newcommand{\kms}{${\rm km\,s}^{-1}$}
\newcommand{\dhi}{D_{\rm HI}}
\newcommand{\magsq}{\mbox{$\rm mag\, arcsec^{-2}$}}

\title{The \hi Mass Function of Galaxies from a Deep Survey in the 21cm Line}
\author{Martin A. Zwaan, Frank H. Briggs, and David Sprayberry}
\affil{Kapteyn Astronomical Institute, Postbus 800, 9700 AV Groningen,
The Netherlands;\\
zwaan@astro.rug.nl, fbriggs@astro.rug.nl, dspray@astro.rug.nl}
\and
\author{Ertu Sorar\altaffilmark{1}}
\affil{Department of Physics \& Astronomy, University of Pittsburgh,
Pittsburgh, PA 15260}

\altaffiltext{1}{current address: 
{\sc Qualcomm} Inc., 6455 Lusk Blvd., San Diego, CA 92121}

\begin{abstract}
 The \hi mass function (H{\sc i}MF) for galaxies in the local universe
is constructed from the results of the Arecibo \hi Strip Survey, a blind
extragalactic survey in the 21cm line.  The survey consists of two
strips covering in total $\sim 65$ square degrees of sky, with a depth
of $cz=7400$~\kms\, and was optimized to detect column densities of
neutral gas $N_{\rm HI}>10^{18} \rm cm^{-2}$ ($5\sigma$).  The survey
yielded 66 significant extragalactic signals of which approximately 50\%
are cataloged galaxies.  No free floating \hi clouds without stars are
found.  VLA follow-up observations of all signals have been used to
obtain better measurements of the positions and fluxes and allow an
alternate determination of the achieved survey sensitivity.  The
resulting H{\sc i}MF has a shallow faint end slope ($\alpha \approx
1.2$), and is consistent with earlier estimates computed for the
population of optically selected gas rich galaxies.  This implies that
there is not a large population of gas rich low luminosity or low surface
brightness galaxies that has gone unnoticed by optical surveys.  The
influence of large scale structure on the determination of the H{\sc
i}MF from the Arecibo \hi Strip Survey is tested by numerical
experiments and was not found to affect the resulting H{\sc i}MF
significantly.  The cosmological mass density of \hi at the present time
determined from the survey, $\ohi(z=0)= (2.0 \pm 0.5) \times
10^{-4}h^{-1}$, is in good agreement with earlier estimates.  We
determine lower limits to the average column densities $\langle N_{\rm
HI} \rangle$ of the galaxies detected in the survey and find that none
of the galaxies have $\langle N_{\rm HI} \rangle < 10^{19.7} \rm
cm^{-2}$, although there are no observational selection criteria against
finding lower density systems.  Eight percent of the signals detected in
the original survey originated in groups of galaxies, whose signals
chanced to coincide in frequency. 
 \end{abstract}

\newpage

\section{Introduction} \label{intro.sect}
 The distribution function of neutral hydrogen masses among galaxies and
intergalactic clouds (the \hi mass function or H{\sc i}MF), and more
generally, the neutral hydrogen density in the nearby universe, $\ohi$,
are important inputs into models of cosmology and galaxy evolution. 
Different attempts have been made to construct an H{\sc i}MF by using
optically selected galaxies (Rao \& Briggs 1993, hereafter RB, Solanes,
Giovanelli \& Haynes 1996).  These studies are based on the assumption
that \hi is always associated with optically bright galaxies.  A major
concern is whether the H{\sc i}MF is complete when it is computed for these
galaxies. 

For example, the population of low surface brightness (LSB) galaxies
might hypothetically constitute a substantial portion of the population
of nearby extragalactic objects (Disney 1976, McGaugh 1996, Dalcanton,
Spergel \& Summers 1997, Sprayberry et al. 1997).  The LSB population
easily escapes detection optically and would not be included in the
samples that are commonly used to evaluate the H{\sc i}MF.  This would
be particularly problematic since LSB galaxies are generally found to be
rich in neutral gas (Schombert et al. 1992, de Blok, McGaugh \& van der
Hulst 1996) and could therefore contribute substantially to the neutral
gas content.  Gas rich dwarf galaxies may also play an important part. 
For example Tyson \& Scalo (1988) have argued that the majority of these
dwarf galaxies remain undetected because only the small portion that is
currently undergoing a rapid phase of star formation is presently
observed with optical telescopes.  A final possible population of gas
rich systems that would escape inclusion in optical catalogs is a class
of intergalactic \hi clouds without stars.  So far, only a few such
systems have been discovered, and they are always found to be
gravitationally bound to a galaxy or a group of galaxies (for example
the Leo ring, Schneider 1989). 

Clearly, the H{\sc i}MF and the \hi content of the local universe should
be measured directly, in such a way that they suffer no bias against gas
rich galaxies or intergalactic \hi clouds which are difficult to detect
optically.  This is possible by means of surveys in the \hi line. 
Several of these surveys have been carried out over the last two
decades. The majority were single dish observations using on/off
techniques, since these surveys were done in conjunction with
observations targeted on cataloged galaxies (Fisher \& Tully 1981b,
Giovanelli \& Haynes 1985, 1989).  Some surveys were concentrated on
groups of galaxies (Haynes \& Roberts 1979, Lo \& Sargent 1979, Fisher
\& Tully 1981a, Hoffman, Lu \& Salpeter 1992) to specifically search
for \hi clouds in the vicinity of known galaxies.  Other surveys were
designed to find \hi signals in voids (Krumm \& Brosch 1984, Szomoru et
al 1996), or to compare voids and superclusters (Weinberg et al. 1991,
Szomoru et al. 1994).  Since these surveys are not pointed at randomly
chosen regions of sky, they may not provide fair tests of the shape of
the H{\sc i}MF or an unbiased measure of the average \hi density.  A few
truly blind surveys have been conducted, the first one in driftscan
mode (Shostak 1977) and one by observing a series of pointings on lines
of constant declination (Kerr \& Henning 1987, Henning 1992).  It is
worrisome that this latter survey could not reproduce the H{\sc i}MF
defined for optically selected galaxies, possibly because the survey was
targeted toward large volumes of known voids.  Another possibility is
that the achieved survey sensitivity is not well understood, leading to
an underestimation of the true number density of \hi rich galaxies and
intergalactic clouds. 

More recently, several surveys have been made using the Arecibo
telescope (Sorar 1994, Spitzak 1996, Schneider 1997).  Surveys are in
progress at Dwingeloo (DOGS) and at Parkes, where the survey will cover
the entire Southern Sky (Staveley-Smith et al. 1996).

In this paper we analyze the Arecibo \hi Strip Survey (Sorar 1994), a
blind survey for extragalactic \hi covering $\sim 65$ square degrees of
sky, out to a redshift of 7400 \kms.  The analysis of the survey results
will concentrate specifically on the understanding of the achieved survey
sensitivity and the vulnerability to large scale structure.  We describe
the details of the survey and the optical and 21cm follow-up
observations in section~\ref{observations.sect}. 
Section~\ref{sensitivity.sect} gives a detailed analysis of the achieved
survey sensitivity.  We present the H{\sc i}MF in section~\ref{himf.sect},
the possible influence of large scale structure on the determination of
the H{\sc i}MF is examined by performing numerical experiments, and the
cosmological mass density of \hi at the present time is calculated in
this section.  In section~\ref{discussion.sect} we compare our findings
with previous estimates of the H{\sc i}MF based on 21cm surveys and optically
selected galaxy samples and discuss the implications of our results. 
Section~\ref{conclusions.sect} summarizes the results.  The distances
used in this paper are based on a Hubble constant $H_0 = 100\, {\rm
km}\,{\rm s}^{-1}\,{\rm Mpc}^{-1}$. 

\section{Observations} \label{observations.sect} 
\subsection{Arecibo \hi Strip Survey (AH{\sc i}SS)}
 The strip survey was carried out on the Arecibo\footnote{The Arecibo
Observatory is part of the National Astronomy and Ionosphere Center,
which is operated by Cornell University under cooperative agreement with
the National Science Foundation.} 305m Telescope in the period of August
1993 until February 1994.  The survey was designed to take advantage of
periods of construction when the telescope pointing was immobilized but
the receiving systems were still operational.  Therefore, the data were
taken in driftscan mode and the telescope beam traced strips of
constant declination.  The same strips were retraced for as many as 30
days in order to obtain very sensitive observations that are capable of
detecting \hi of low surface density.  The main survey was divided in
two sessions: one survey covering 10.5 hours of RA at $\delta =
14^{\circ}14'$, and the second survey covering 9.7 hours of RA at
$\delta = 23^{\circ}09'$.  All observations were made at night.  The
limiting column density was $10^{18} \mbox{cm}^{-2}$ (at a $5\sigma$
level) for gas filling the telescope beam, which subtends 3 $h^{-1}$ kpc
at 3 $h^{-1}$ Mpc and 70 $h^{-1}$ kpc at 74 $h^{-1}$ Mpc.  The total sky
coverage was approximately 65 square degrees, with a depth of
$cz=7400$~\kms.  The survey was capable of detecting \hi masses of
$6\times 10^5 h^{-2} \msol$ at 7 $h^{-1}$ Mpc and $1.5\times 10^8 h^{-2}
\msol$ at the full depth of the survey, in the main beam which has a
FWHM of 3 arcmin.  The first sidelobe of the telescope beam pattern
could detect $1.5\times 10^9 h^{-2} \msol$ at the full depth of the
survey, which makes the sidelobes effective in detecting high \hi mass
galaxies over a 15 arcmin wide strip.  The details of the AH{\sc i}SS
and the data reduction are described by Sorar (1994).  A summary of the
reduction path is described by Briggs et al.  (1997). 

The survey yielded a total of 61 detections, of which approximately half
could be associated with cataloged galaxies listed in the NASA
Extragalactic Database (NED\footnote{The NASA/IPAC Extragalactic
Database (NED) is operated by the Jet Propulsion Laboratory, California
Institute of Technology, under contract with the National Aeronautics
and Space Administration.}).  Five detections have no obvious
counterparts on the Digitized Sky Survey (DSS\footnote{Based on
photographic data of the National Geographic Society -- Palomar
Observatory Sky Survey (NGS-POSS) obtained using the Oschin Telescope on
Palomar Mountain.  The NGS-POSS was funded by a grant from the National
Geographic Society to the California Institute of Technology.  The
plates were processed into the present compressed digital form with
their permission.  The Digitized Sky Survey was produced at the Space
Telescope Science Institute under US Government grant NAG W-2166.  })
although they are more than $10^\circ$ away from the Galactic plane. 

\begin{figure}[htb]
\epsfxsize=11.5cm 
\epsfbox{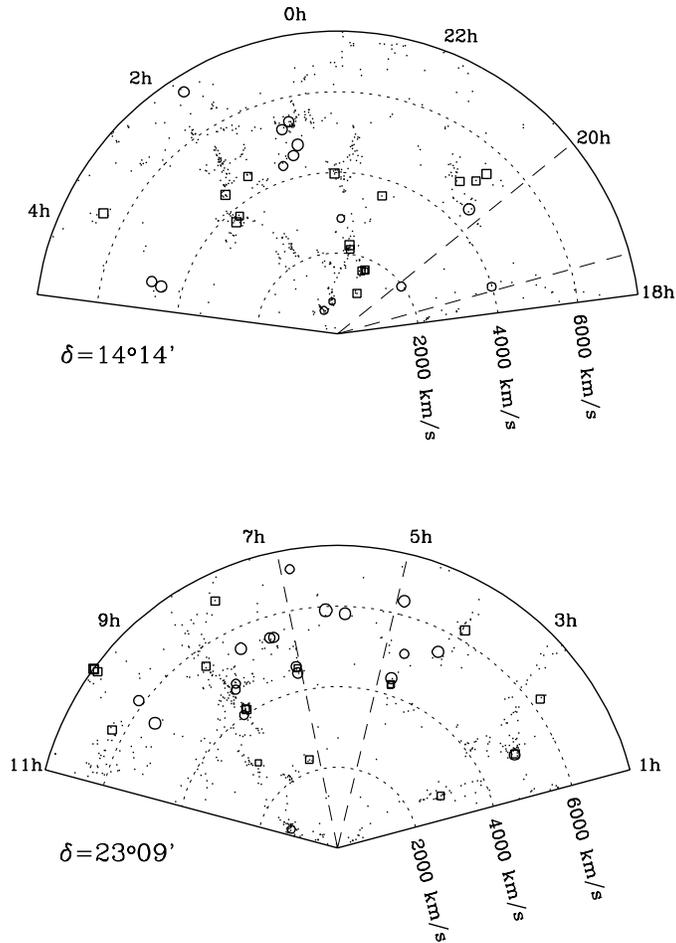}
\caption{\small  Slice diagrams of the location of all the \hi
selected objects from the Arecibo \hi Strip Survey.  Boxes indicate
uncataloged galaxies, circles indicate cataloged galaxies.  The size of
the symbols reflects the \hi mass of the galaxies.  Also shown are
galaxies from the CfA catalog (Geller \& Huchra 1989) within a
$10^{\circ}$ strip centered on the declination of the surveys.  The
Galactic plane intersects the strips at RA$\approx 6^{\rm h}$ and
RA$\approx 19^{\rm h}$ as indicated by the dashed lines. 
 \label{slices.fig}}
\end{figure}

Fig.~\ref{slices.fig} shows slice diagrams of the location of all \hi
selected objects compared with galaxies in the CfA catalog (Geller \&
Huchra 1989).  Galaxies within a $10^{\circ}$ strip centered on the
declination of the strip are plotted.  Different symbols are used to
distinguish between cataloged (circles) and uncataloged (boxes)
galaxies.  The size of the symbols reflects the \hi mass of the
galaxies, in the sense that bigger symbols indicate larger \hi masses. 
The Galactic plane intersects the strips at $\rm RA \approx 6^h$ and
$\rm RA \approx 19^h$ as indicated by the dashed lines.  The CfA catalog
is obviously incomplete in these regions of the sky due to Galactic
extinction, whereas the \hi survey suffers no bias against detection in
these regions.  It is clear from this figure that \hi selected galaxies
generally follow the same structures as the optical galaxies.  In
particular, if we only consider those regions of the sky where the CfA
catalog is complete, we find that more than 80\% of the \hi selected
galaxies lie in regions where the average galaxies density is higher
than the mean density.  This is consistent with the finding that LSB
galaxies and gas-rich dwarfs lie on structures delineated by normal,
high surface brightness galaxies (Bothun et al.  1986, Thuan, Gott \&
Schneider 1987, Eder et al.  1989, Thuan et al.  1991, Mo, McGaugh \&
Bothun 1994). 
 Furthermore, none of the \hi selected galaxies in the sections
where the CfA is complete are found in regions where the galaxy density
is less than one fifth of the cosmic mean.  This is in agreement with
the results of Szomoru et al.  (1994) and Weinberg et al.  (1991) that
\hi selected galaxies are not found in selected void fields.
A more detailed analysis of the large scale distribution of the \hi
selected galaxies will be presented elsewhere.

\subsection{21cm Follow-up observations}
 Follow-up 21cm synthesis observations on all signals found in this
survey have been performed with the NRAO\footnote{The National Radio
Astronomy Observatory is a facility of the National Science Foundation
operated under cooperative agreement by Associated Universities, Inc.}
Very Large Array (VLA) in D-configuration.  These follow-up observations
are essential for three reasons:\\
 1) As stated before, the survey was capable of detecting signals as far
as 7 arcmin from the center of the survey strip.  Consequently, the
coordinates yielded by the survey have uncertainties on the order of
several arcminutes.  The association with a cataloged galaxy or a galaxy
on the DSS is therefore not always unambiguous, especially in those
cases where there are several prominent galaxies visible within a 15
arcmin wide strip.  21cm observations with spatial resolution of $\sim
1$ arcmin, are sufficient to obtain unique identifications.\\
 2) Flux measurements from the AH{\sc i}SS can be poor if a signal is
detected at large distance from the center of the beam.  In principle, a
correction to the flux could be made since the response function of the
telescope is known with reasonable accuracy, but this is only possible
if the positional accuracy is sufficient.  Furthermore, measurements
from the survey spectra might underestimate the integral flux if the
source is more extended than the primary beam.  Thus, a rough measure of
the \hi distribution increases confidence in the analysis.  \\
 3) Some signals can be caused by pairs or small groups of galaxies,
whose line emission might stack up in the same channels.  It is not
obvious from the Arecibo survey spectra which signals are caused by more
than one galaxy.  In fact, this situation was found by the VLA
observations to occur in five cases. 
 
Short VLA observations ($\sim$ 20 min) of all 61 detected galaxies were
performed during the D-configuration sessions in May 1995 and September
1996.  The signal of three systems fell below the detection limit of the
snap-shot observation.  These systems were re-observed in the
D-configuration during the second session, but with longer integration
times ($\sim$ 3 hours), resulting finally in confirmation of all 61
detections at levels consistent with the AH{\sc i}SS sensitivity.  The three
weaker signals originated in galaxies whose declinations are at the
center of the AH{\sc i}SS. 

The VLA observations were performed using 63 channels over a 3.125 MHz
bandwidth, corresponding to a velocity range of approximately 660 \kms. 
On-line Hanning smoothing was used, resulting in a velocity resolution
of $\sim 10.5$ \kms.  Phase calibrators were observed once for each
source.  Only seven different phase calibrators were used.  Each time
that the phase calibrator was changed, a primary flux calibrator (3C48
or 3C286) was observed in order to tie the flux scales together. 
Therefore, the primary calibrators were not observed with the same
correlator settings as each galaxy, but only with six correlator
settings.  The overhead time due to slewing of the telescopes and
observations of primary calibrators is significantly decreased this way. 
This technique provided adequate passband calibration for these short
observations, where high dynamical range or accurate channel-to-channel
flux density calibrations are not needed.  There were no strong
continuum sources in these fields. 

Since the observing conditions were generally quite good, little editing
was necessary to remove interference and bad baselines.  Continuum was
removed from the data in the $uv$-plane, by making linear fits to the
real and imaginary components for each visibility in the line-free
channels and subtracting the appropriate values from all channels.  The
$uv$ data were calibrated and transformed to datacubes using natural
weighting.  Using natural weighting rather than uniform weighting
results in a slightly lower spatial resolution, but higher sensitivity. 
The resolution in the transformed datacubes is $\sim 60'' \times 60''$. 
The final r.m.s noise was approximately 1.2 mJy/beam for each channel
(corresponding to a minimal detectable column density of $\sim 1.7
\times 10^{19} \mbox{cm}^{-2}$ [$5\sigma$]) for the short integrations
and 0.8 mJy/beam (limiting column density $\sim 9 \times 10^{18}
\mbox{cm}^{-2}$ [$5\sigma$]) for the longer integrations.  Since the
observed galaxies were generally barely resolved, {\sc clean}ing of the
data was not really necessary.  Nonetheless, because the synthesized
beam of these short observations has strong deviations from a Gaussian
shape, we chose to {\sc clean} the datacubes and restore them with a
Gaussian beam in order to make an effective search for companion objects
throughout the primary beam. 

Total \hi masses were calculated using $M_{\rm HI}/\msol = 236 \,d^2 \int
SdV$, where $d$ is the distance to the source in Mpc, and $S$ is the
flux density in mJy over profile width $\Delta V$ in \kms.  Total \hi
maps were constructed by adding the regions in the channel maps that
contain line emission.  Contour maps of \hi emission and global profiles
will be presented in a future paper. 

\begin{figure}[htb]
\epsfxsize=12cm 
\epsfbox{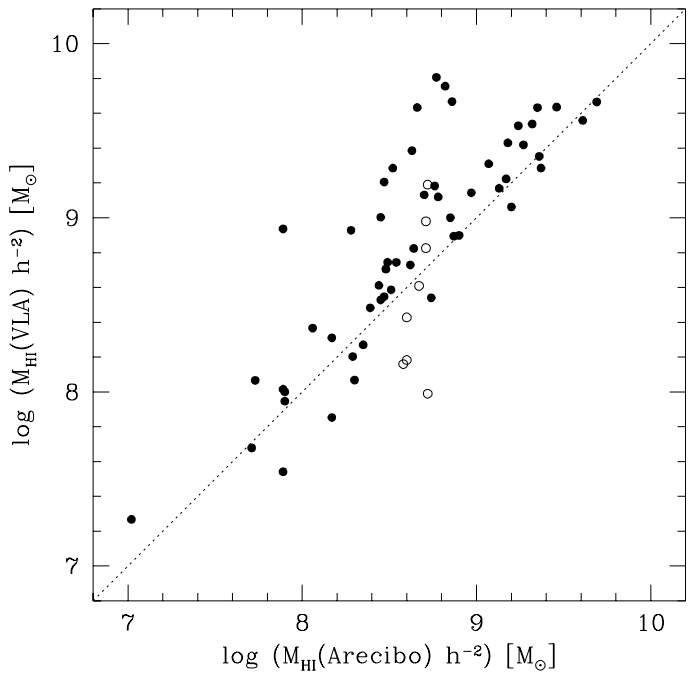}
\caption{\small  Comparison of total \hi masses as measured with
the VLA and \hi masses derived from the Arecibo \hi Strip Survey
spectra.  The dashed line is the line of equality.  Different symbols
distinguish between singe detections (filled circles) and multiple
detections (open circles).  \label{compmhi.fig}}
\end{figure}

Fig.~\ref{compmhi.fig} shows the \hi masses as measured with the VLA
plotted against the \hi masses that are derived from the AH{\sc i}SS spectra. 
Different symbols are used to distinguish between single detections
(filled circles) and multiple detections (open circles).  Arecibo
measurements of these multiple detections are of course always
unreliable estimates of the real \hi masses.  This figure clearly
illustrates the fact that the \hi masses derived from the Arecibo
spectra generally underestimate the true \hi masses, and that follow-up
observations were essential to determine accurate measurements.  In the
analysis of the AH{\sc i}SS results we will use \hi masses calculated from
the VLA observations, except for galaxies with integrated VLA fluxes
less than $1.0~{\rm Jy\,km\,s}^{-1}$.  Galaxies with these low fluxes
are only found at small distances from the center of the Arecibo beam,
where the normalized response function is close to unity.  Arecibo
measurements for these galaxies are therefore reliable estimates of
their real fluxes. 

Table~\ref{tabel} summarizes the global parameters of the detected
galaxies which are derived from the VLA observations.  The following
information is contained: Column 1: Identification number of the Arecibo
detection.  Indices indicate multiple detections.  Column 2: Name of
galaxy if already cataloged.  Columns 3 and 4: B1950 coordinates. 
Column 5: Logarithm of \hi mass.  Column 6: Heliocentric velocity,
calculated by taking the mean of the velocities at 20\% of peak flux
density.  Column 7: Declination offset from center of the survey strip. 
Column 8: Identification code for multiple detections.

 \subsection{Optical follow-up observations} We have also started a
program of optical $B$-band imaging of the \hi selected galaxies with
the 2.5m Isaac Newton Telescope\footnote{The Isaac Newton Telescope is
operated by the Royal Greenwich Observatory in the Spanish Observatorio
del Roque de los Muchachos of the Instituto Astrofisica de Canarias.}
(INT) on La Palma.  During four observing runs in the period between
October 1995 and March 1997 we have been able to obtain images of all
target galaxies.  The first noteworthy result of the follow-up optical
observations is that {\it all} of the \hi sources more than $10^\circ$
away from the Galactic plane are found to be associated with
galaxies in the $B$-band images.  We have seen no indication that any of
the sources detected in the Arecibo survey are anything other than
`ordinary' galaxies having both gas and stars.  Put another way, we have
so far failed to find any \hi sources that are pure \hi clouds without
stars.  The analysis of the optical observations will be presented in a
subsequent paper. 

\section{The survey sensitivity} \label{sensitivity.sect}
 The improved coordinates and \hi fluxes derived from the VLA
observations help to verify the completeness limit of the survey.  The
detectability is determined in the following way:

The \hi mass in a detected signal can be expressed as $\mhi/\msol = 236 \times
S \, \Delta V d^2$, where $d$ is the distance to the source in Mpc, and
$S$ is the flux density in mJy that here is considered to be constant
over a rectangular profile of width $\Delta V$ km/s.  The sensitivity of
an observation is optimal when the spectra are smoothed to the velocity
width of the source.  With optimal smoothing the noise becomes
$\sigma(\Delta V)=\sigma_0 \sqrt{\Delta V_0/\Delta V}$, where $\Delta
V_0$ is the spectral resolution of the receiving system and $\sigma_0$
is the noise level in the unsmoothed spectra.  The limiting flux density
$S_{\rm c}$ for a $5 \sigma$ detection is then given by $S_{\rm
c}(\Delta V) = 5 \sigma(\Delta V) = 5 \sigma_0 \sqrt{\Delta V_0/\Delta
V}$. For the AH{\sc i}SS, the average noise level after coadding spectra
taken at different days was 0.75 mJy for a velocity resolution of 16
km/s. 
 
The normalized response function of the survey telescope, $I(\theta)$,
describes the relative response to a source which is detected at an
offset $\theta$ from the center of the survey strip.  In other words,
$I(\theta)$ is the integral of the flux density sensed by the telescope
as a source makes a cut through the beam pattern, missing the center of
the beam by angle $\theta$, normalized in such a way that $I(0)=1$.  For
the Arecibo telescope this function falls off to $\sim 0.125$ at
$\theta=3.25'$ and reaches a second maximum of $\sim 0.2$ at
$\theta=4.8'$ due to the high sidelobe level.  We define a correction
factor $c_{\rm r} (\theta)$ that accounts for the shape of the response
function by $c_{\rm r} (\theta) = 1/I(\theta)$.  The limiting flux
density of a galaxy at an offset $\theta$ from the center of the survey
strip can then be expressed as $S_{\rm c}c_{\rm r}(\theta)$.  In
general, an \hi source should be detected by the survey if its flux
density exceeds this limiting flux density, that is $S > S_{\rm c}c_{\rm
r}(\theta)$.  This can be rewritten as
 \begin{equation} \label{detect.eq}
 D \equiv \frac{\int S\,dV \sqrt(\Delta V_0/\Delta V)}
 {5\sigma_0 \Delta V_0 c_{\rm f}} > c_{\rm r}(\theta),
 \end{equation}
 where we define $D$ as the detectability.  The factor $c_{\rm f}$
represents the normalized feed gain of the telescope, which is a
function of frequency.  $c_{\rm f}$ can be approximated by an analytical
expression of the form $c_{\rm f} \approx 1-([f - f_0]/w)^2$, where $f$
is the frequency in MHz, $f_0$ is the center of the survey band and $w$
is a parameter which determines the width of the band.  During the
observations at $\delta=14^\circ 14'$ the shape of the normalized feed
gain remained unchanged and could be satisfactorily fit by
$f_0=1395$~MHz and $w=40.5$~MHz.  During the $\delta=23^\circ 09'$
observations, the gain was retuned a few times to the settings:
$f_0=1395$~MHz and $w=37.5$~MHz and $f_0=1410$~MHz and $w=33.8$~MHz. 
The $c_{\rm f}$ dependence of $D$ leads to a `distance dependence' for
the flux density sensitivity. 

\begin{figure}[htb]
\epsfxsize=12cm 
\epsfbox{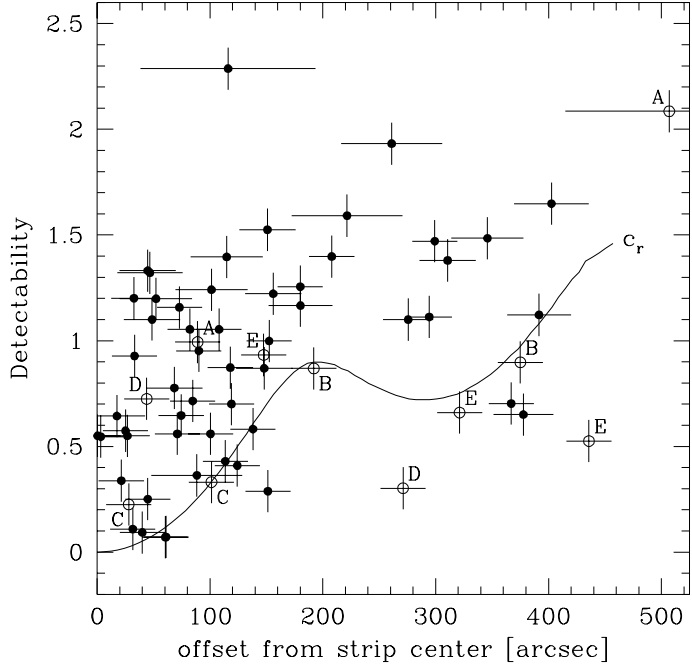}
\caption{\small  Detectability, $D$, of the objects versus the
offset from the center of the survey strip.  Multiple detections are
indicated by open symbols and coded to indicate which galaxies originate
from the same Arecibo detection.  The solid curve shows the detection
limit.  Objects below this line should be undetectable if their \hi
profile widths would be rectangular.  \label{sens.fig}}
\end{figure}

The sensitivity can now be verified by plotting the detectability $D$ of
the objects against declination offset from the center of the survey
strip.  This is shown in Fig.~\ref{sens.fig}.  All signals that were
detected are shown in this plot.  The horizontal errorbars are the
result of a combination of positional accuracy in the VLA maps and the
spatial extent of the galaxies in the direction orthogonal to the survey
strip.  The signals corresponding to pair or group detections are
indicated by open circles and letters A to E are used to identify these
related or confused signals in Table~\ref{tabel}.  Filled circles mark
all single object detections.  The solid line represents $c_{\rm
r}(\theta)$, the limit to the detectability.  In principle, all filled
circles should lie above the solid line, the area below this line is
`undetectable'.  The line is a satisfactory limit to the data points,
especially if we consider the naive character of Eq.\ref{detect.eq}. 
That is, this equation assumes that the detected profiles are symmetric
and featureless.  Since the profiles are generally heavily smoothed,
this is a reasonable assumption, but lop-sided profiles or strong double
horned profiles might exceed the detection limit while they are formally
`undetectable' according to Eq.\ref{detect.eq}. 

The limiting depth $d_{\rm c}$, the maximum distance to which the object
could be placed and still remain within the sample, can now be expressed
as a function of $\mhi$, $\theta$ and $\Delta V$:
 \begin{equation}
 d_{\rm c}(\mhi,\theta,\Delta V)= \sqrt{\frac{\mhi \, I(\theta)}{ 236\, S_{\rm c}(\Delta V)
 \, \Delta V}}. 
 \end{equation}
 
The variation of feed gain with frequency imposes a minor correction to
the limiting depth $d_{\rm c}$ which can be compensated by solving for
$d_{\rm c}$ in \begin{equation} S(d_{\rm c}) c_{\rm f}(d_{\rm c}) =
S_{\rm c}, \end{equation} that is, the true flux density $S(d_{\rm c})$
of a galaxy located at the distance limit $d_{\rm c}$ multiplied by the
feed gain at a frequency corresponding to that distance has to be equal
to the limiting flux density $S_{\rm c}$ obtained at the center
frequency, where the feed gain is optimized and the nominal noise level
of the system is normally calculated. 

The next step is to calculate the effective search volume of the survey. 
The volume $d{\cal V}$ of a slice of the total survey volume at
declination $\delta$ and with length $l$ (the total length of the strips
in radians, $l=\Delta {\rm RA}\, 2\pi \cos \delta / \rm 24^h$) and width
$d\theta$ is given by
 \begin{equation}
 d{\cal V}(\mhi,\theta,\Delta V) = \left\{
 \begin{array}{ll}
  \frac{1}{3} \, d_{\rm c}^3(\mhi,\theta,\Delta V) \, l \,d\theta, 
                           & {\rm if} \,d_{\rm c} < d_{\rm BW},\\
  \frac{1}{3} \, d_{\rm BW}^3 \, l \,d\theta, 
                           & {\rm if} \,d_{\rm c} > d_{\rm BW},
 \end{array} \right. 
\end{equation}
 where $d_{\rm BW}$ is the limiting depth of the survey imposed by the
bandwidth of the receiving system.  The total survey volume that is
sensitive to a galaxy with \hi mass $\mhi$ and velocity spread $\Delta
V$ can then be calculated by taking the integral over $\theta$.  For
galaxies in the flux limited regime, for which $d_{\rm c} < d_{\rm BW}$
for each $\theta$, this integral simply becomes:
 \begin{equation} \label{volume1.eq}
 {\cal V}(\mhi,\Delta V) =
 \frac{1}{3} \int_{-\infty}^{\infty} d_{\rm c}^3 l \, d\theta = 
 \frac{1}{3} \int_{-\infty}^{\infty} \left( \frac{\mhi\,I(\theta)}{236 
 \, S_{\rm c}\,\Delta V} \right)^{3/2} l \, d\theta. 
 \end{equation}
 For galaxies in the bandwidth limited regime, $d_{\rm c} > d_{\rm BW}$
for $\theta$ smaller than a certain critical value $\theta_{\rm BW}$. 
The integral must now be split up in separate parts for the flux limited
regimes and the band width limited regime:
 \begin{eqnarray} \label{volume2.eq}
 {\cal V}(\mhi,\Delta V) &=&
 \frac{1}{3} \int_{-\infty}^{-\theta_{\rm BW}} d_{\rm c}^3 l \, d\theta +
 \frac{1}{3} \int_{-\theta_{\rm BW}}^{\theta_{\rm BW}} d_{\rm BW}^3 l \, d\theta +
 \frac{1}{3} \int_{\theta_{\rm BW}}^{\infty} d_{\rm c}^3 l \, d\theta \nonumber\\
 &=& \frac{2}{3} \int_{0}^{\theta_{\rm BW}} d_{\rm BW}^3 l \, d\theta +
 \frac{2}{3} \int_{\theta_{\rm BW}}^{\infty} \left( \frac{\mhi\,I(\theta)}{236 
 \, S_{\rm c}\,\Delta V} \right)^{3/2} l \, d\theta,
 \end{eqnarray}
 where we made use of the symmetry of the beam shape in the last step. 

This effective search volume is still dependent on two variables: $\mhi$
and $\Delta V$.  The most convenient parameterization of $\cal V$ for
computing an H{\sc i}MF is to express $\cal V$ as a function of $\mhi$
only.  This can be achieved by adopting a relation between $\mhi$ and
$\Delta V$.  Such a relation is known to exist since optical luminosity
$L$ is related to $\Delta V$ via the Tully-Fisher relation (Tully \&
Fisher 1977), and $L$ is related to $\mhi$ as $\mhi \propto L^{0.9}$
(see Briggs 1990).  Briggs \& Rao (1993, hereafter BR) determined the
$\mhi$-$\Delta V$ relation empirically by plotting $\Delta V$ against
log$\mhi$ for 1139 optically selected galaxies from the catalog by
Fisher \& Tully (1981b).  A fit to these points gives $\Delta V = 0.16
\, \mhi^{1/3}$.  Recently, Salpeter \& Hoffman (1996) analyzed \hi
observations of 70 dwarf galaxies and find a similar trend: $\Delta V
\propto \mhi^{0.36}$.  This relation is therefore valid over a wide
range in \hi mass.  Our data are also in good agreement with this
relation.  Note that the velocity widths in these relations are not the
inclination corrected maximum rotational velocities, but just the
observed velocity spreads.  

\begin{figure}[htb]
\epsfxsize=14cm 
\epsfbox{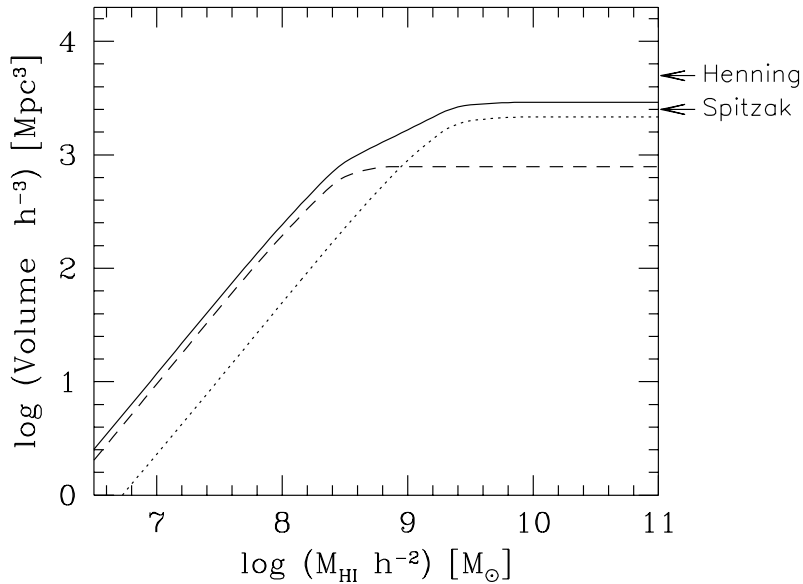}
\caption{\small  The effective search volume vs.  \hi mass.  The
solid line is the total search volume of the Arecibo \hi Strip
Survey, the dashed and the dotted line show the effective search volume
corresponding to the main beam and the sidelobes, respectively.  The
arrows indicate the survey volumes of two \hi surveys which
are comparable in size to the AH{\sc i}SS: Henning (1995) and Spitzak
(1996). \label{volume.fig}}
\end{figure}

The effective survey volume ${\cal V}(\mhi)$ can now be calculated by
substituting the relation between $\Delta V$ and $\mhi$ found by BR and
$S_{\rm c}(\Delta V)$ into Eq.\ref{volume1.eq} and Eq.\ref{volume2.eq}. 
This volume as a function of $\mhi$ is shown by the solid line in 
Fig.~\ref{volume.fig}.  In the flux limited regime, for \hi masses $<
10^{8.5}$, the search volume is ${\cal V} \propto \mhi^{5/4}$.  (A
proportionality often used in the literature is ${\cal V} \propto
\mhi^{3/2}$ [e.g.,  Henning 1995, Schneider 1997].  This power 3/2 arises
if the dependence of $\Delta V$ on $\mhi$ is discarded, effectively
assuming that all galaxies have the same profile width.) In the high
mass region ($\mhi> 10^{9.5}$), the limiting depths are no longer
determined by the detectability of the signals, but simply by the
bandwidth of the receiving system.  Therefore, the effective survey
volume for high mass systems is not dependent on \hi mass or $\Delta V$. 
The total survey volume in this regime, is $\sim 3000 h^{-3} {\rm
Mpc}^3$.  For lower \hi masses the volume decreases rapidly and is only
$\sim 1.0 h^{-3} {\rm Mpc}^3$ for $\mhi=10^7 \msol$.  The dashed and the
dotted line show the effective search volume corresponding to the main
beam and the sidelobes, respectively.  This figure clearly shows that
the sidelobes do not add much volume in the low mass range, but are very
effective in finding large \hi masses. 


 Since the observed velocity width is related to inclination $i$ and
maximum rotational velocity $V_{\rm max}$ as $\Delta V = 2 V_{\rm max}
\sin i$, the limiting depth is $d_{\rm c} \propto (\Delta V)^{-1/4}
\propto (\sin i)^{-1/4}$.  The limiting depth is therefore weakly
dependent on inclination, in the sense that low inclined galaxies can be
detected to slightly larger distances.  Consequently, an \hi survey
would in principle preferentially select face-on galaxies.  In practice,
this effect is negligible, since the expectation for $\langle (\sin
i)^{-1/4} \rangle$ of a randomly oriented sample is close to unity:
\begin{equation}
\langle (\sin i)^{-1/4} \rangle = 
\frac{\int_0^{\pi/2}(\sin i)^{-1/4} \sin i\, d i}
     {\int_0^{\pi/2} \sin i\, d i} \approx 1.086. 
 \end{equation}
 Only the flux limited regime would be hampered by the inclination
effect, but most galaxies in this regime are low mass, dwarf galaxies. 
The velocity spreads are small for these galaxies, and turbulent motions
play an important part at establishing the profile width (cf.,  Lo,
Sargent \& Young 1993). 

A potential hazard in radio surveys is the influence of radio frequency
interference (RFI).  RFI can in principle cause false negatives (miss a
significant signal) in the sample if it affects the spectrum exactly at
the frequency of a source, or when a narrow-lined source is mistaken for
RFI.  However, the driftscan method that has been used for the AH{\sc
i}SS has proven to provide a very good stability for RFI signals (see
Briggs et al.  1997).  Repetitive coverage of the same regions of sky
makes the survey immune to RFI and unstable baselines. This
has been demonstrated by the fact that all signals identified in the
Arecibo survey have been confirmed by the VLA follow-up.

\section{The \hi mass function} \label{himf.sect}
 The \hi mass function (H{\sc i}MF) is defined analogously to optical
luminosity functions.  The H{\sc i}MF $\Theta(\mhi/\mhis) d(\mhi/\mhis)$
gives the total number of galaxies or intergalactic clouds per Mpc$^3$
in the mass interval $d(\mhi/\mhis)$ centered on $\mhi/\mhis$.  Here, we
find it convenient in our analysis and figures to plot the H{\sc i}MF as
the number of galaxies or intergalactic clouds per decade in mass.
In order to parameterize the shape of the H{\sc i}MF, we
adopt the conventional Schechter (1976) function,
 \begin{equation} \label{himf.eq}
 \Theta(\frac{\mhi}{\mhis}) d(\frac{\mhi}{\mhis})
 = \theta^*(\frac{\mhi}{\mhis})^{-\alpha} \exp-(\frac{\mhi}{\mhis}) d (\frac{\mhi}{\mhis}),
 \end{equation}
 with free parameters $\alpha$, the slope of the low-mass end, $\mhis$
the characteristic mass that the defines the kink in the function and
$\theta^*$, a normalization factor.  This function must be integrated
over decade bins for comparison with the binned data in the figures.

\subsection{Methods}
 The classical method (see e.g.,  Christensen 1975, Schechter 1976) for
determining luminosity functions is based on the assumption that
galaxies are distributed in a uniform manner.  The luminosity function
is determined by dividing the number of galaxies in a bin centered on
$M$ by ${\cal V}(M)$, the effective search volume for that particular
$M$.  This method is easily applicable to the AH{\sc i}SS.  The search volume
can be evaluated with Eq.\ref{volume1.eq} and Eq.\ref{volume2.eq}, using
a statistical relation between velocity width and \hi mass.  The
advantages of this method are that it is nonparametric and that it is
automatically normalized.  The important disadvantage is that it assumes
homogeneity and its use might lead to errors in $\Theta(\mhi)$ if
density fluctuations due to large scale structure occur on distance
scales comparable to, or greater than the depth $d_{\rm c}$ at which
$\mhi$ can be detected. 

A slightly modified form of the `classical' method is the $\sum (1/{\cal
V}_{\rm max})$ method, first used by Schmidt (1968), which is also
applicable to \hi surveys.  Instead of calculating a mean survey volume
for each mass bin, this method consists of summing the reciprocals of
the volumes corresponding to the maximum distances to which the objects
could be placed and still remain within the sample.  The values of
${\cal V}_{\rm max}$ can be calculated directly using
Eq.\ref{volume1.eq} and Eq.\ref{volume2.eq}, using now the measured
velocity spread instead of the statistical value.  The two procedures
give similar results when the number of galaxies per bin is large.  For
the less densely populated bins the two procedures can give different
results because ${\cal V}_{\rm max}$ of a particular galaxy can strongly
deviate from the average $\cal V$ of the mass bin it falls in.  In
section~\ref{results.sect} the H{\sc i}MF is calculated using Schmidt's method. 
Like the classical method, Schmidt's method is also vulnerable to errors
caused by large scale structure.  The possible effects of large scale
structure are discussed in section~\ref{lss.sect}. 

\subsection{Results} \label{results.sect}
\begin{figure}[htb]
\epsfxsize=12cm
\epsfbox{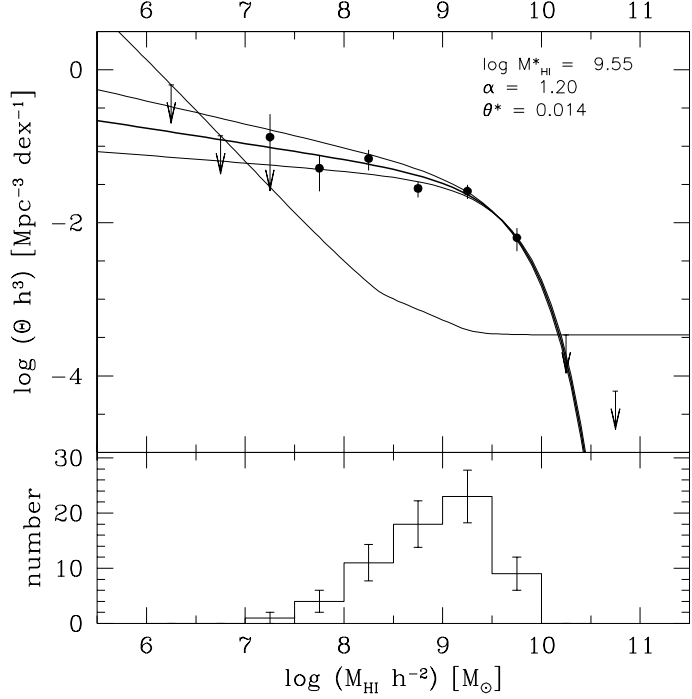}
\caption{\small  Lower panel: The distribution of H~{\sc i} masses of
the detected galaxies from VLA follow-up measurements.  The errorbars
are given by Poisson statistics.  Upper panel: The thin line is the
sensitivity of the survey defined by $\phi=1/{\cal V}$, where $\cal V$
is the effective search volume.  In the region $10^7 {\rm M}_\odot < M_{\rm HI} <
10^{10} {\rm M}_\odot$ this function defines an upper limit to the space density
of intergalactic H~{\sc i} clouds without stars.  The measured H~{\sc i} mass
function per decade is shown by the points.  The fat line is a Schechter
luminosity function with parameters as given in the upper right corner. 
Also Schechter functions with $\alpha=1.1$ and $\alpha=1.3$ are shown. 
The arrows show upper limits to the space density of H~{\sc i} rich
galaxies or intergalactic H~{\sc i} clouds.  The two arrows on the right
are from a complementary survey with the Arecibo telescope over the
range 19,000 to 28,000 ${\rm km\,s}^{-1}$.  \label{himfhist.fig}}
\end{figure}

 Fig.~\ref{himfhist.fig} shows the principal results of this analysis. 
The lower panel shows the observed distribution of \hi masses binned per
half decade, with errorbars given by Poisson statistics.  This histogram
shows that the survey has detected galaxies with \hi masses in the range
from $10^7$ to $10^{10} \msol$ and therefore enables us to determine the
H{\sc i}MF over three orders of magnitudes in \hi mass. 

The solid dots in the upper panel of Fig.~\ref{himfhist.fig} show the
H{\sc i}MF determined by the $\sum (1/{\cal V}_{\rm max})$ method.  The
errorbars are given by Poisson statistics.  Also drawn in this figure
are analytical curves given by the Schechter function of
Eq~\ref{himf.eq}.  A satisfactory fit to the points is obtained with
$\alpha = 1.20$, $\theta^*=0.014~ \rm Mpc^{-3}$ and $\log (\mhis/\msol)
= 9.55$.  Mass functions with faint end slopes of 1.10 and 1.30 are
shown to indicate the uncertainty in the value of $\alpha$.
 We note that in the present analysis the parameterization of the H{\sc
i}MF in the form of a Schechter function is only used to enable
comparison with other \hi survey results and results based on the
optically selected galaxy population.  The $\sum (1/{\cal V}_{\rm max})$
method recovers the shape and amplitude of the H{\sc i}MF simultaneously
without using a Schechter function (or any other parameterization) as an
assumption about the intrinsic shape of the H{\sc i}MF. 


The observational limits to the determination of the H{\sc i}MF are also
illustrated in Fig.~\ref{himfhist.fig} by the thin line.  This line
represents the sensitivity function of the survey to objects of \hi mass
$\mhi$, defined as $\phi(\mhi)=1/{\cal V}(\mhi)$, where ${\cal V}(\mhi)$
is the effective search volume.  In the range $10^7 \msol < \mhi <
10^{10} \msol$ this function defines an upper limit to the space density
of intergalactic \hi clouds without stars.  It also shows that this
survey is not capable of measuring the H{\sc i}MF directly in the regions
$\mhi < 10^7 \msol$ and $\mhi > 10^{10} \msol$ if the extrapolation of
the analytical form for $\Theta(\mhi)$ holds.  The sensitivity function
does allow us to define upper limits in these ranges as indicated by
arrows.  The upper limit at $\mhi = 10^{10.75} \msol$ is from the
complementary survey with the Arecibo telescope over the range 19,000 to
28,000 \kms, only sensitive to high mass galaxies (See Sorar (1994) for
details). 

\subsection{Influence of large scale structure} \label{lss.sect}
 The slice diagrams in Fig~\ref{slices.fig} show that the survey strips
sample a wide range of large scale structure, as the combined RA range
extends nearly 2/3 of the way around the sky.  On the other hand, the
limited depth of the survey for small \hi masses might cause the low
mass end of the H{\sc i}MF to be affected by local density fluctuations. 
Therefore, a major concern in the determination of the mass function
following Schmidt's method, is whether homogeneity is a fair assumption. 
For instance, a local density enhancement would overestimate the number
of low mass galaxies or clouds and would give rise to an overestimate of
the faint end slope of the mass function.  The biases in the shape of
the mass function due to large scale structure can be avoided by making
use of a maximum likelihood estimator (cf.,  Efstathiou, Ellis \&
Peterson 1988, Saunders et al. 1990). If the spatial density distribution
of galaxies and intergalactic clouds is given by $N(\mhi,\vec r)$, then
the overall density $\rho(\vec r)$ and the mass function $\Theta(\mhi)$
can be separated as $N(\mhi,\vec r)=\rho(\vec r)\Theta(\mhi)$.  Maximum
likelihood estimators can then be used to solve either for
$\Theta(\mhi)$ without knowledge of $\rho(\vec r)$ or solve for
$\rho(\vec r)$ without knowledge of $\Theta(\mhi)$ . 

Although the maximum likelihood method is a powerful procedure for
determining optical luminosity functions (Marzke, Geller \& Huchra 1994,
Lin et al. 1996, Saunders et al. 1990), we chose not to apply it to the
AH{\sc i}SS.  The most important problem is that we have to deal with
small number statistics.  Especially the faint end slope of the mass
function is defined by very few galaxies or clouds per bin.  Some
experimentation with application of the algorithm to small samples such
as ours showed that maximum likelihood methods can produce erratic
results is these situations.  Another complication arises because
maximum likelihood methods assume that the shape of the \hi mass
function is independent of the space density $\rho(\vec r)$.  In
practice, high mass galaxies will have a higher statistical weight in
the determination of $\rho(\vec r)$ since they are simply more numerous
in our sample.  It is questionable whether this space density defined by
the high mass galaxies is a fair estimate for the galaxies at the faint
end side of the mass function, in the region that is dominated by dwarf
and LSB galaxies.  Although these galaxies are found to follow the same
general large scale structures as the normal HSB galaxies, they
preferentially avoid the highest density regions (Mo et al. 1994, Taylor
1996).  In principle, the spatial density $\rho$ could be separated into
different functions for individual morphological types or different
ranges in \hi mass, but the limited number of galaxies in our sample
does not allow this differentiation. 

In order to investigate the effect of large scale structure we have
performed numerical experiments. These tests consisted of randomly
placing artificial galaxies in a volume with weights given by the
H{\sc i}MF given in Eq.\ref{himf.eq}. 
A range of H{\sc i}MF parameters were investigated. The results are
illustrated using the derived H{\sc i}MF parameters from
Fig~\ref{himfhist.fig}:  $\theta=0.014~ \rm Mpc^{-3}$, $\log
(\mhis/\msol) = 9.55$ and $\alpha=1.20$. The galaxies have random
inclination, and rotational velocity related to \hi mass as determined
by BR. Galaxies were selected from these volumes in the same manner as
the Arecibo \hi Strip Survey selects galaxies from the sky.


\begin{figure}[htb]
\epsfxsize=12cm
\epsfbox{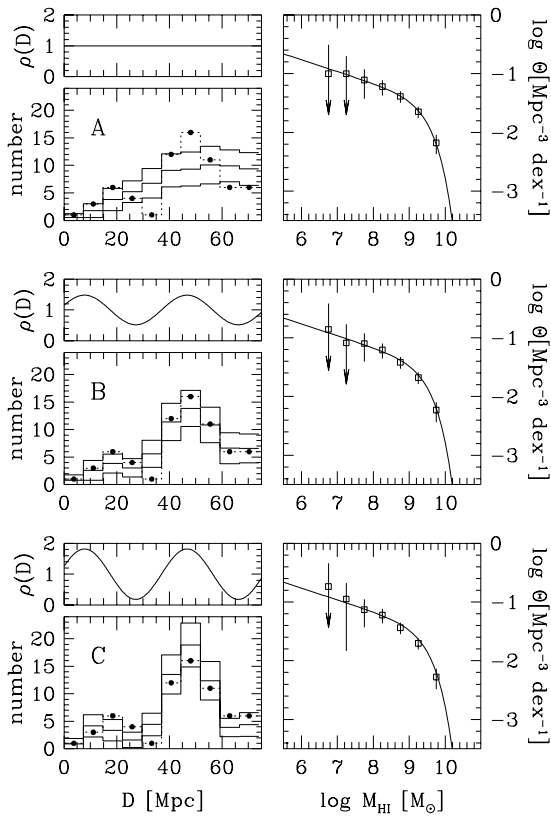}
\caption{\small  Numerical tests of the influence of large scale
structure on the determination of the H{\sc i}MF.  The left panels show
the imposed normalized density fluctuations in the artificial catalogs. 
The histograms show the redshift distributions of the simulated data as
solid lines, with $1\sigma$ uncertainties as thin lines.  The measured
redshift distribution of our survey sample is represented by solid
circles and a dashed line.  The right panels show the $\sum (1/{\cal
V}_{\rm max})$ H{\sc i}MF of the simulated data as open squares, with $1
\sigma$ errorbars.  The input H{\sc i}MF is indicated by the solid
curves.  \label{mosaic.fig}}
\end{figure}

Fig.~\ref{mosaic.fig} shows the results of three experiments, labeled as
A, B and C.  The presented results are each the average of 100
independent simulations.  The top panels show the assumed normalized
density fluctuations $\rho$.  The central panels show the redshift
distribution of the simulated data as thick-lined histograms, thin solid
lines show $1\sigma$ uncertainties.  Overlaid on these histograms is the
measured redshift distribution of our survey sample, indicated by solid
circles and dashed histograms.  The bottom panels show the averaged
H{\sc i}MFs of the simulated catalogs as open squares, with $1\sigma$
errorbars computed as the standard deviations of the 100 simulations. 
The $\sum 1/{\cal V}_{\rm max}$ method has been used. 

In case A a homogeneous density distribution is assumed.  It is not
surprising to see that the input H{\sc i}MF and the measured H{\sc i}MF
agree with high accuracy.  The $\sum 1/{\cal V}_{\rm max}$ method should
give reliable results in this situation.  It is however apparent that
the survey observations have several distance bins that are overdense or
underdense compared to the uncertainty band of the `uniform density'
simulation.  The comparison implies that we may be observing
underdensities at $D \approx 30$ Mpc and $D \approx 70$ Mpc and an
overdensity at $D \approx 50$ Mpc.  To test this hypothesis we
constructed catalogs with a density fluctuation as indicated by the top
panel in case B.  The precise functional form of this fluctuation is not
important, the condition is that is should produce consecutive under and
overdensities at $D \approx$ 30, 50 and 70 Mpc.  The central panel shows
that the imposed density fluctuation reproduces the observed redshift
distribution satisfactorily.  It is surprising however that if the $\sum
1/{\cal V}_{\rm max}$ method is applied to these simulated data the
resulting H{\sc i}MF is still indistinguishable from the input H{\sc
i}MF.  In other words, the $\sum 1/{\cal V}_{\rm max}$ method appears to
be a robust method, and not very sensitive to the effects of large scale
structure.  This is even true for case C, where a more severe density
fluctuation is imposed, and where the resulting H{\sc i}MF is still in
good agreement with the input H{\sc i}MF.  We can conclude from these
simulations that although we see a hint of the effects of large scale
structure in our data, the observed deviations from uniformity have no
significant influence on our determination of the H{\sc i}MF. 

\section{Discussion} \label{discussion.sect}
\subsection{Previous estimates of the H{\sc i}MFs from \hi surveys}
\label{comphi.sect}
 The first \hi surveys could only be used to set upper limits to the
space density of intergalactic \hi clouds without stars, and did not
yield enough detections to allow the determination of the shape of the
H{\sc i}MF.  Shostak (1977) was to first to define these limits by means
of blind driftscan emission and absorption searches in the 21cm line
with the NRAO 91m telescope.  Due to the relatively poor sensitivity (12
- 40 mJy) and the small total effective search volume ($\sim 300 h^{-3}
\rm Mpc^{3}$), these limits were not very strict.  In the region
$\mhi<10^{8.5}\msol$ the limits set by the Arecibo \hi Strip Survey are
at least two orders of magnitude lower (99\% confidence) than those set
by Shostak.  Haynes \& Roberts (1979), Lo \& Sargent (1979) and Fisher
\& Tully (1981a) search for invisible galaxies in groups of galaxies. 
Haynes \& Roberts conclude from observations in the direction of the
Sculptor group that intergalactic \hi clouds with $\mhi > 10^8 \msol$ do
not exist.  Lo \& Sargent find four previously uncataloged LSB dwarf
galaxies but their upper limits to the space density of unseen objects
do not improve Shostak's.  Fisher \& Tully used the NRAO 91m
telescope to search for invisible galaxies in the M81 group.  Their null
result allowed them to push the upper limits to the space density 0.6
dex lower those set by Shostak.  Also Krumm \& Brosch (1984) find no \hi
sources in their driftscan searches of void regions and are only able
to define upper limits.  These limits are not very strict since their
survey is only sensitive in the redshift range above 5300 \kms, and is
consequently only capable of finding \hi masses $>10^{10} \msol$. 

\begin{figure}[htb]
\epsfxsize=12cm
\epsfbox{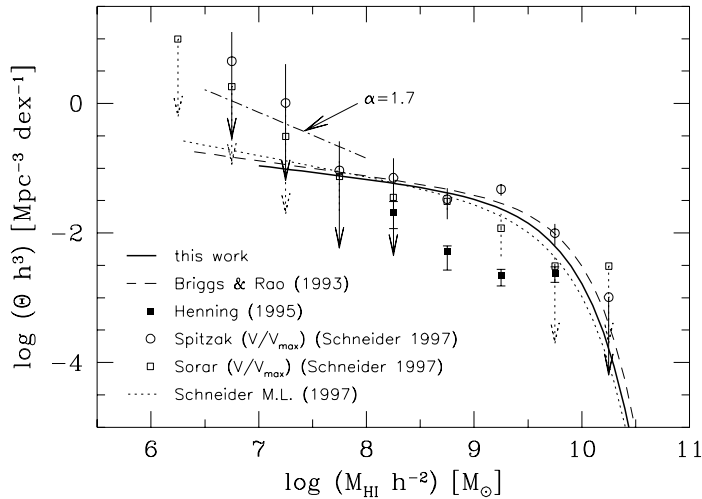}
\caption{\small  Comparison of \hi mass functions of different
surveys.  The fat solid line shows the H{\sc i}MF presented in this
paper, using VLA observations of galaxies found in the Arecibo \hi Strip
Survey.  Briggs \& Rao's (1993) H{\sc i}MF based on the Fisher \& Tully
catalog is indicated by dashed curve.  The solid squares are from
Henning (1995).  The open symbols result from Schneider's (1997) ${\cal
V}/{\cal V}_{\rm max}$ method applied to the data of Spitzak (1995)
(circles) and the uncorrected Arecibo data from Sorar (1994) (squares). 
Schneider's (1997) maximum likelihood solution is indicated by the
dotted curve.  The dash-dot line shows a faint end slope of
$\alpha=1.7$.  \label{compa.fig} }
\end{figure}

A series of papers by Kerr \& Henning (1987) and Henning (1992, 1995)
describes blind surveys which consisted of observing a series of
pointings along lines of constant declination in the zone of avoidance
and at high galactic latitude.  These surveys yielded 39 detections (of
which half were previously unknown) and were the first to put serious
constraints on the shape of the H{\sc i}MF.  For comparison, the H{\sc
i}MF calculated by Henning (1995) is reproduced in Fig.~\ref{compa.fig}
as solid squares, together with the AH{\sc i}SS H{\sc i}MF represented
as a solid line.  Henning's points are significantly lower than our
measured curve.  According to Henning's calculations objects with
$10^{8.5} \msol < \mhi < 10^{9.5} \msol$ are deficient by a one order of
magnitude compared to the AH{\sc i}SS H{\sc i}MF.  The presence of the
Local Void in Henning's survey could explain a part of this discrepancy,
but even if the void is omitted the points are still beneath our H{\sc
i}MF.  The most reasonable explanation for the discrepancy is that the
sensitivity of Henning's survey is not well understood, leading to
a serious underestimate of the H{\sc i}MF. 


The most recent determination of the H{\sc i}MF is the one by Schneider
(1997) who used both the uncorrected Arecibo data from Sorar's (1994)
survey and the data from Spitzak's (1996) survey with the Arecibo
telescope.  Schneider assessed the completeness limits of the surveys
with the use of ${\cal V}/{\cal V}_{\rm max}$ tests.  He calculated the
average value of ${\cal V}/{\cal V}_{\rm max}$ and scaled that to the
correct value of 0.5 by increasing the limiting flux density $S_{\rm
c}$.  Schneider noted the vulnerability of this method to the influence
of large scale structure.  Despite this different approach, the
resulting H{\sc i}MF is in good agreement with our estimate, as can be
seen in Fig.~\ref{compa.fig}.  Only the low mass end deviates
significantly.  The statistics in the bins with $\mhi \leq 7.25$ are
poor, and Schneider could not rule out a slope as high as $\alpha=1.7$. 
The differences in the slopes can be partly explained by the fact that
{\it uncorrected} \hi masses are used in Schneider's estimate of the
H{\sc i}MF from Sorar's data.  The effect that the flux is
underestimated when the Arecibo beam misses a galaxy by a certain angle
$\theta$ was not taken into account.  The VLA \hi masses that are used
in our H{\sc i}MF are on average 50\% higher, leading to a shift of
galaxies to higher \hi mass bins.  As an alternate approach, Schneider
also applied a maximum likelihood fit to all the signals brighter that
$20\sigma$.  The resulting H{\sc i}MF had a faint end slope
$\alpha=1.32$.  This result is also drawn in Fig.~\ref{compa.fig} and is
in very good agreement with our function over the whole range in \hi
mass. 

In addition to making comparisons with surveys that are equally blind as
the AH{\sc i}SS, it is also interesting to compare our results with
surveys that cover comparable volumes in underdense and overdense
regions.  It is noteworthy that the H{\sc i}MF determined from H{\sc i}
observations in clusters (McMahon 1993), as well as that determined in
cosmic voids (Szomoru et al.  1996) are both well fit by a Schechter
function, with approximately the same shape as as our H{\sc i}MF.  The
scaling $\theta^*$, however, differs by a factor of ten or more between
these different regions. 

Since the AH{\sc i}SS crosses the Galactic plane twice, and consequently
a large fraction of the total survey volume lies in regions which are
not complete in optical catalogs, it is difficult to calculate the exact
void filling factor of our survey.  If we limit the calculation to the
regions where $|b|>30^\circ$, we estimate that approximately 50\% of the
survey volume consists of regions where the average galaxy density is
less than one third of the cosmic mean, and approximately 10\% of the
volume consists of structures where the galaxy density takes on more
than three times the cosmic mean.  The same percentages are obtained if
random slices through the Universe are taken.  If voids would be omitted
from the survey, the scaling $\theta^*$ of the H{\sc i}MF would increase
by a factor of two.

\subsection{Comparison with H{\sc i}MFs bases on optically selected galaxies}
\label{compopt.sect}
 The construction of an H{\sc i}MFs from a sample of optically selected
galaxies can either be done statistically, by studying optical
luminosity functions and the dependence of \hi mass on on optical
luminosity, or directly by using 21cm data of optically selected
galaxies.  The first method was used by RB, who reviewed the literature
that describes luminosity functions and \hi richness for individual
morphological types.  Combining these data enabled them to construct an
H{\sc i}MF over three orders of magnitude in \hi mass.  Their result is in
good agreement with our estimate of the H{\sc i}MF in the range of \hi mass
where our survey is sensitive. 

The second method is discussed by BR who analyzed \hi observations drawn
from the catalogs of Fisher \& Tully (1981b) and Hoffman et al. (1989),
and recently by Solanes et al. (1996) who use 21cm data of an optical
magnitude-limited sample of galaxies in the direction of the
Pisces-Perseus supercluster.  Both authors arrive at the same values for
the normalization and $\mhis$, but their values of the faint end slope
differ significantly.  BR find $\alpha \sim 1.25$ while Solanes et al.
find a declining slope.  The Solanes et al. result may not be relevant to
the discussion here since their sample excluded all dwarf irregular
galaxies and contained no galaxies with $\mhi < 10^{8.5} \sim
\frac{1}{10}\mhis$.  The H{\sc i}MF determined by BR is reasonably
consistent with with our estimate of the H{\sc i}MF as shown in
Fig.~\ref{compa.fig}. 

\subsection{Implications: A new \hi selected galaxy population?}
 The comparison between our result and the H{\sc i}MFs based on
optically selected galaxy samples provides a direct test of the
existence of a new population of gas rich galaxies that has gone
unnoticed by optical surveys.  In a way, this comparison appraises the
completeness of optically selected catalogs for gas rich galaxies.  Any
contribution of uncataloged gas rich dwarf or gas rich LSB galaxies
would yield an difference between H{\sc i}MFs computed from optically
selected and HI-selected galaxy catalogs. Fig~\ref{compa.fig} shows that
the H{\sc i}MF derived from the results of the AH{\sc i}SS is in very
good agreement with previous estimates based on optically selected      
galaxy catalogs.  This implies that the optically selected samples that 
have been used to evaluate the H{\sc i}MF are remarkably complete.
There is no evidence for a large number of neutral gas rich objects that
have escaped inclusion in these catalogs.  To the extent that optically
selected catalogs are incomplete for LSB galaxies, the excluded galaxies
must be predominantly gas poor, consistent with the finding by
Sprayberry et al. (1997) that these excluded LSB galaxies are
predominantly of low optical luminosity.

Several authors have speculated about the existence of a large class of
gas rich dwarf galaxies (e.g.,  Dekel \& Silk 1986, Tyson \& Scalo 1988). 
These galaxies would yield a steep rise in the H{\sc i}MF below $\mhi
\approx 10^8 \msol$.  We find no observational support for the existence
of this class of galaxies.  Based on the counting statistics in the mass
bins $10^{6.5}$ to $10^8 \msol$ we can exclude a faint end slope of
$\alpha=1.7$ or steeper with a 99\% confidence level. 

\subsection{Neutral gas density} \label{neutral.sect}
 Our knowledge of $\og$ at high redshift is determined by the statistics
of Damped Ly-$\alpha$ systems, seen in absorption against background
quasars (e.g.,  Lanzetta, Wolfe \& Turnshek 1995, Storrie-Lombardi, Irwin
\& McMahon 1996).  The picture emerging from these studies is that $\og$
reaches a maximum at $z \approx 3$ and has been declining since then. 
At low redshifts different effects complicate the determination of
$\og$.  Firstly, at redshifts $z < 1.6$, the Ly-$\alpha$ line is not
redshifted to optical wavelengths and has to be observed from space
(Lanzetta et al. 1995, Rao, Turnshek \& Briggs 1995).  Secondly, the
evaluation of $\og$ may depend on the selection effects in the sample of
quasars that has been used.  Especially at these low redshifts, it is
very difficult to compile a unbiased sample of quasars since QSOs would
need to be observed within the optical images of galaxies.  At the one
hand, gravitational lensing can bring faint quasars into the sample
which should otherwise be below the detection limit (Smette, Claeskens
\& Surdej 1997).  But on the other hand, dust in Ly-$\alpha$ systems
might obscure the background objects to a level where they are
undetected (Fall \& Pei 1993).

The Arecibo \hi Strip Survey can be used to evaluate the gas density
differently.  The mentioned problems are circumvented since $\og$ is not
measured by using the Ly-$\alpha$ line, but the 21cm line in emission. 

\begin{figure}[htb]
\epsfxsize=12cm
\epsfbox{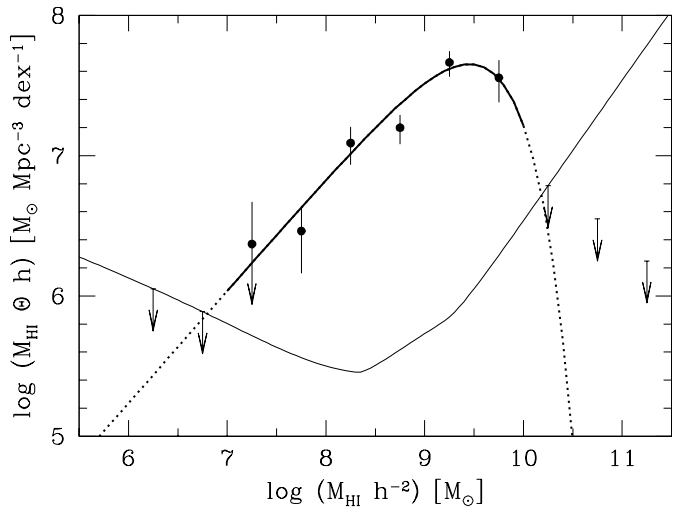}
\caption{\small  Space density of \hi mass contained in objects
of different masses per decade.  The fat line shows the converted
analytical H~{\sc i} mass function calculated by multiplying $\Theta
\times \mhi$, where $\Theta$ is the H{\sc i}MF plotted in
Fig.~\ref{himfhist.fig}.  The dashed regions of the curve indicate that
the sample contains no galaxies with $\mhi < 10^7 \msol$ or $\mhi >
10^{10} \msol$.  The thin line indicates the sensitivity of the survey. 
The arrows mark upper limits to the space density of H~{\sc i} mass. 
The three arrows on the right are from a complementary survey with the
Arecibo telescope over the range 19,000 to 28,000 \kms.  Note that
galaxies with $10^9 \msol < \mhi < 10^{10} \msol$ contribute most to the
integral H~{\sc i} density.  \label{hidens.fig}}
\end{figure}

 The space density of \hi contained in objects of different \hi masses
is plotted in Fig.~\ref{hidens.fig}.  The solid line indicates an
analytical expression of this function, determined from the product
$\mhi \Theta(\mhi)$.  The thin line represents the sensitivity limits
and the arrows mark upper limits determined analogously to the upper
limits in Fig~\ref{himfhist.fig}. 


The integral \hi mass density at the present epoch can be determined by
taking the integral over the solid line in Fig.~\ref{hidens.fig}.  This
yields: $\rhi = \Gamma(2-\alpha) \theta^* \mhis$, where $\Gamma$ is the
Euler gamma function.  Using the best fit Schechter parameters, we
derive that $\rhi(z=0) = 5.8~\times~10^7\,h \,{\rm M}_\odot~\mbox{Mpc}^{-3}$
or $3.9~\times~10^{-33}\,h\, \mbox{g\,cm}^{-3}$, with a statistical
error of $20\%$.  A summation over all survey galaxies
$\rhi=\sum\mhi/{\cal V}_{\rm max}$ yields a slightly smaller value:
$\rhi(z=0) = 5.4~\times~10^7\,h \,{\rm M}_\odot~\mbox{Mpc}^{-3}$ since this
calculation does not include the contribution to $\rhi$ of the density
function below $\mhi=10^7 \msol$ or above $\mhi=10^{10} \msol$.  The
cosmological mass density of \hi at $z=0$ is $\ohi(z=0)= (2.1 \pm 0.4)
\times 10^{-4}h^{-1}$.  The total cosmological mass density of neutral
gas at the present epoch is $\og(z=0) = (2.7 \pm 0.5) \times
10^{-4}h^{-1}$, assuming that the mass percentage of \hei is 25\% of the
total gas density. 

This value agrees surprisingly well with earlier estimates by RB, who find
$\rhi(z=0) = (4.8 \pm 1.1 )~\times~10^7\,h \,{\rm M}_\odot~\mbox{Mpc}^{-3}$ by
using optically selected galaxies.  Fall \& Pei (1993) arrive at
approximately the same value of $\ohi$ by simply computing the average
$\mhi/L_B$ of the Huchtmeier \& Richter (1989) catalog and multiply that
by the mean luminosity density in the local universe as estimated by
Efstathiou et al. (1988).  The agreement between the cosmological mass
density based on optically selected galaxies and that based on \hi
selected galaxies illustrates once more that there is not much neutral
gas hidden in objects like LSB galaxies, dwarfs or intergalactic \hi
clouds that are missed by optical surveys. 

The contribution of dwarf galaxies to the \hi content of the nearby
universe is modest, galaxies with \hi masses $<10^8 \msol$ make up only
$\sim 17\%$ of the integral \hi density.  The density function in
Fig.~\ref{hidens.fig} clearly illustrates that the integral \hi mass
density is dominated by high mass galaxies with \hi masses in the range
$10^9 \msol < \mhi < 10^{10} \msol$, which are $L^*$ galaxies.  At $\mhi
\sim 10^{10} \msol$ the density function drops off sharply, indicating
that Malin 1 type galaxies make no significant contribution to $\ohi$. 
This sharp cutoff was already noted by Bothun (1985).  Much stronger
upper limits to the contribution of galaxies with $\mhi > 10^{10} \msol$
will be set in the near future by the results of the Parkes Multibeam
Survey (Staveley-Smith et al.  1996). 

The estimate of the integral gas density from the AH{\sc i}SS is a robust
result.  Two effects cause a relatively low uncertainty in the
determination of $\ohi$.  Firstly, the peak in the \hi gas density
function is conveniently caused by the galaxies that dominate the
counting statistics.  Galaxies in the lower mass bins, where the Poisson
errors are large, contribute not much to the total density and therefore
also not much to the uncertainty in $\ohi$.  Secondly, the effective
search volume in the mass region that dominates the density function is
mostly band width limited.  Uncertainties in $\mhi$ and velocity width
do not influence the search volume in this regime.

\subsection{What could be missed?}
 The question arises which gas rich systems could be missed by the
survey and could still contribute significantly to the integral gas mass
density in the local universe.  The only possible candidates that might
escape detection are extremely low gas density systems with \hi column
densities below $10^{18} \rm cm^{-2}$, the detection limit of the
AH{\sc i}SS.  

It has been shown that the gas density of spiral galaxies is
correlated with the optical surface brightness in such a way that lower
optical surface brightness implies lower gas densities.  De Blok et al.
(1996) observed a sample of LSB galaxies and showed that the neutral gas
densities are generally a factor of two lower than those of late type
high surface brightness galaxies.  Furthermore, van der Hulst et al.
(1993) has shown that the gas densities of LSB galaxies are often just
below or equal to the critical density for star formation.  If the
relation between optical surface brightness and gas density extends to
still lower surface brightnesses this implies that galaxies that most
easily escape detection in optical surveys are also the ones that might
be missed in 21cm surveys. 

Until recently, blind \hi surveys were not able to reach column
densities much lower than $10^{20} \rm cm^{-2}$.  These surveys would
therefore still miss low density systems, and would not be capable to
set strict constraints on the number density of LSB galaxies.  However,
the AH{\sc i}SS is sensitive to column densities $\sim 10^{18} \rm
cm^{-2}$ at a $5\sigma$ level for gas filling the beam.  Using an
extrapolation of the scaling of Disney \& Banks (1997), $N_{\rm HI} =
10^{20} (M_{\rm HI}/L_B) 10^{0.4(27-\mu_B)}$, where $\mu_B$ is the
surface brightness averaged over the \hi disk, and a typical value of
$M_{\rm HI}/L_B = 1$ for LSB galaxies, the AH{\sc i}SS would be capable
of finding galaxies with $\mu_B=32~\magsq$.  Even if a ten times lower
value of $M_{\rm HI}/L_B$ is used, and if the area of the galaxy were
ten times smaller than the area covered by the Arecibo beam, still
galaxies as faint as $\mu_B=27~\magsq$ would be detectable. 

The VLA observations of the AH{\sc i}SS galaxies can be used to set lower
limits to average column densities for the sample.  Due to the low
spatial resolution of the observations ($\sim 1'$), the measured column
density is in many cases an average $\langle N_{\rm HI} \rangle$ over
the entire projected surface of the \hi layer of the galaxy.  If the
galaxy is spatially unresolved, an upper limit to the extent $D_{\rm
HI}$ of the \hi layer can be determined, leading to a lower limit of the
average column density $\langle N_{\rm HI} \rangle \propto \mhi D_{\rm
HI}^{-2}$.  Although many of the galaxies in the sample are unresolved,
and many of the true values of $\langle N_{\rm HI} \rangle$ may be
higher, we find no values of $\langle N_{\rm HI} \rangle$ below
$10^{19.7}\rm cm^{-2}$.  Hence, all galaxies in the sample have normal
\hi column densities, even though there are no observational selection
effects against finding extreme low density systems.  There is no
indication of the existence of a group of extreme low column density
galaxies that has been missed by previous \hi surveys, simply because
they were not capable of reaching the same low column densities as the
AH{\sc i}SS.  
Because we seem to be observing a lower limit to $\langle
N_{\rm HI} \rangle$, much higher than our detection limit, it is very 
unlikely that galaxies with even lower column densities are missed by   
the AH{\sc i}SS.

Theoretical predictions of the ionization of gas layers by the
extragalactic UV background (Corbelli \& Salpeter 1993, Maloney 1993,
Charlton, Salpeter \& Linder 1994) demonstrate physical mechanisms that
can explain the possible non-existence of low column density neutral gas
layers.  These models predict photoionization by the extragalactic UV
background of low column density regions, below $10^{19.5}\rm cm^{-2}$. 
Furthermore, models by Quinn, Katz \& Efstathiou (1996) show that the
ionization only moderately suppresses the formation of galaxies with
rotational speeds larger than 50 \kms, but that it seriously affects the
low density regions around these systems.  The models are verified by
very deep VLA observations on one galaxy which appears to have a sharp
truncation of the HI disk below a column density of $10^{19.5}\rm
cm^{-2}$ (van Gorkom et al.  1993).  Further confirmation of ionization
of low density HI comes from recent observations by Bland-Hawthorn et
al.  (1997) who have detected ionized gas beyond the truncated HI disk
in NGC 253. 

\subsection{\hi self absorption}
 The calculation of the total \hi masses in this paper is based on the
assumption that the optical depth of the \hi layer is close to zero. 
Any possible influence of \hi self absorption, which will cause an
underestimation of the true \hi mass, is ignored.  In this paragraph we
will make a rough estimate of the influence of \hi self absorption on
the determination of the cosmological mass density $\ohi$ and the H{\sc i}MF. 
The possible effect of self absorption could apply to all H{\sc i}MFs
compared in this paper (see Section~\ref{comphi.sect},
\ref{compopt.sect} and Fig.~\ref{compa.fig}), since none of these have
addressed this possibility.

The problem of self absorption for galaxies can be assessed statistically
by plotting the 21cm flux of different Hubble types as a function of
inclination $i$ to the line of sight.  The line of sight through a
inclined galaxy will be larger, generally causing a higher fraction of
self absorption.  Haynes \& Giovanelli (1984) use data of a sample of
1500 galaxies with 21cm fluxes measured with Arecibo.  They define a
correction factor $f_{\rm HI}$ which is defined by the corrected flux
divided by the measured flux and find a general trend: $f_{\rm HI} =
(\cos i)^{-c}$ where $c$ is a constant dependent on morphological type. 
The values of $c$ are found to be 0.04 for Sa and Sab, 0.16 for Sb and
0.14 for Sbc and Sc galaxies.  No correlations are found for
morphological types earlier than Sa or later than Sc, indicating that
self absorption is negligible in these types.  Higher self absorption in
types Sb to Sc can be explained by the fact that these galaxies
generally have the highest \hi surface densities (Cayatte et al. 1994). 
Furthermore, rotation curves of early type spiral galaxies show the
strongest deviations from solid body rotation implying large velocity
spreads per line of sight, which further decreases self absorption. 
 
Mean self absorption factors per morphological type can be obtained by
averaging $f_{\rm HI}$ over a random distribution of inclinations.  This
yields 
\begin{equation}
\langle f_{\rm HI} \rangle 
= \frac{\int_0^{\pi/2} (\cos i)^{-c} \sin i\,d i}
       {\int_0^{\pi/2} \sin i\, d i}
= 1/(1-c).  
\end{equation}
The mean correction factors then become
1.04 for Sa and Sab, 1.19 for Sb and 1.16 for Sbc and Sc galaxies. 

The cosmological \hi mass density is dominated by high mass galaxies,
which are statistically most likely to be late type spirals.  The
correction to $\ohi$ due to \hi self absorption will therefore probably
not be more than the value of $\langle f_{\rm HI} \rangle$ averaged over
morphological types Sb to Sd.  Assuming that all types Sb to Sd
contribute equally to $\ohi$, we derive that the mean value of $\langle
f_{\rm HI} \rangle$ is 1.10.  Even in the most pessimistic approach, the
correction factor can not be more than 1.19. 

What will be the effect of \hi self absorption on the shape of the H{\sc
i}MF? Using the same arguments as above, we conclude that the effect on
the high mass range will be marginal.  The normalization factor
$\theta^*$ and the value which determines the kink, $\mhis$, will
probably increase by no more than 10\%.  Galaxies that determine the
faint end slope $\alpha$ are mostly dwarf and LSB galaxies.  On the one
hand, self absorption may be unimportant because the gas density in
these galaxies is usually low (van der Hulst 1993) and the dust content
is presumably low (McGaugh 1994) which implies scarcity of clumped gas
(Haynes \& Giovanelli 1984).  On the other hand, the rotation curves of
dwarf and LSB galaxies often show solid body rotation (de Blok et al. 
1996, Swaters 1997) which causes a low velocity spread along a line of
sight, leading to high fractions of self absorption.  These two
counteracting effects make the value of $\langle f_{\rm HI} \rangle$ for
low mass galaxies uncertain, but probably higher than that for high mass
galaxies.  After applying the correction factor to the low mass
galaxies, some galaxies will shift to higher mass bins, eventually
leading to a slightly shallower faint end slope of the mass function. 
The conclusion that the faint end slope of the mass function is shallow
will therefore not be altered by the effects of \hi self absorption. 

\section{Conclusions} \label{conclusions.sect}
 We have used the Arecibo \hi Strip Survey in combination with 21cm
follow-up observations with the VLA to determine the \hi mass function
of gas rich galaxies in the local universe.  The resulting H{\sc i}MF is in
good agreement with earlier estimates based on samples of optically
selected galaxies.  This implies that there is not a large population of
gas rich dwarfs or low surface brightness galaxies, previously
undetected by optical surveys.  The observed faint end slope of the
H{\sc i}MF is flat ($\alpha \sim 1.2$) leaving no room for a large class of
gas rich dwarfs.  The cosmological mass density of \hi in the local
universe is $\ohi(z=0)= (2.0 \pm 0.5) \times 10^{-4}h^{-1}$, also
consistent with earlier estimates.  The neutral gas content is dominated
by high mass galaxies with $10^9 \msol < \mhi < 10^{10} \msol$. 
The observations indicate the existence of a lower limit to the average
\hi column density of 19.7 $\rm cm^{-2}$, consistent with theoretical
predictions concerning the ionizing extragalactic UV background. 

\acknowledgments
 We thank 
G. Bothun,
E. de Blok, 
P. Sackett,  
A. Szomoru, 
M. Verheijen,
and the referee J. van Gorkom for useful comments.

\begin{table}
\dummytable\label{tabel}
\end{table}

\begin{figure}
\plotfiddle{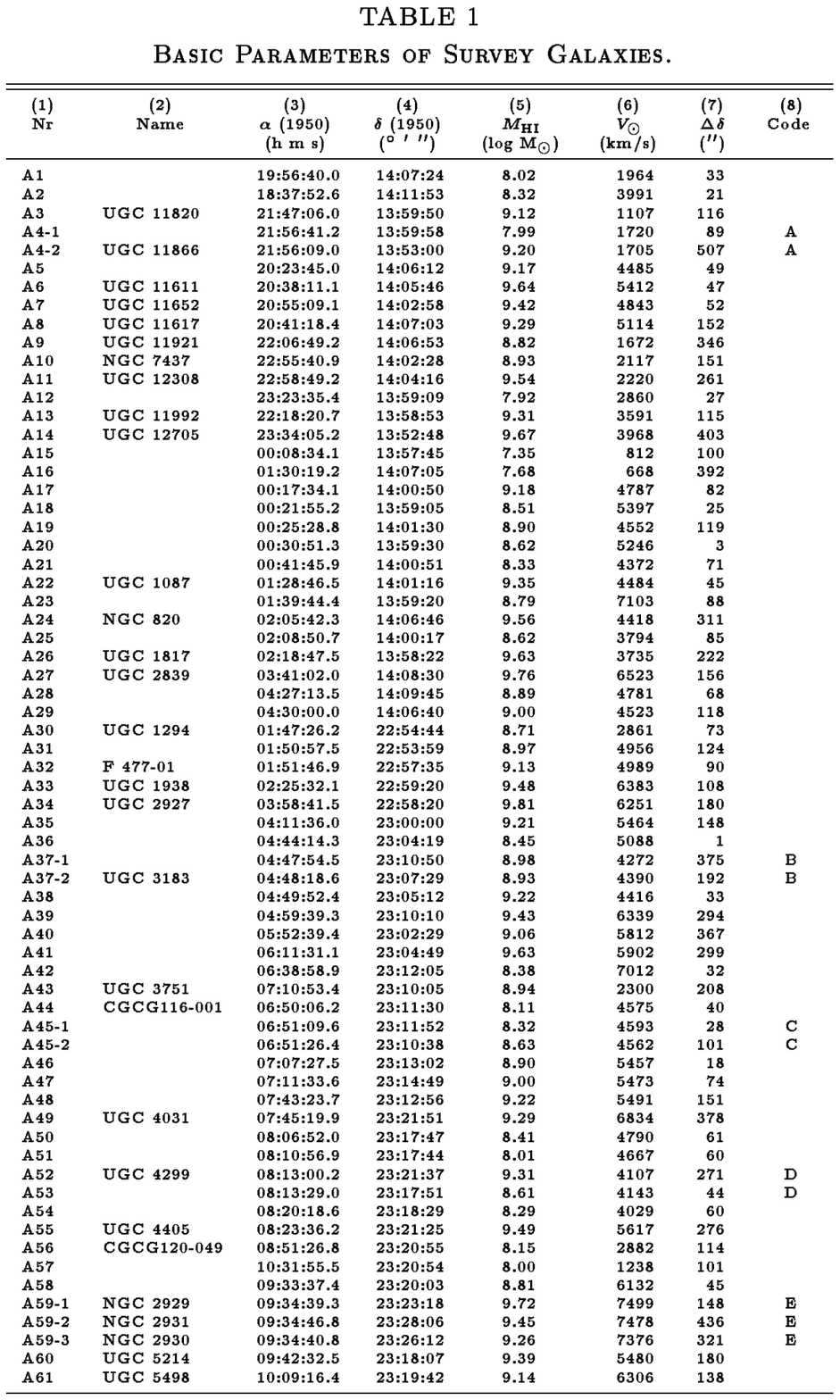}{10cm}{0}{118}{118}{-357}{-372}
\end{figure}

\end{document}